\documentclass[aps,pra,twocolumn,showpacs,superscriptaddress]{revtex4-1} 
\usepackage[colorlinks=true,linkcolor=blue,citecolor=blue,urlcolor=blue]{hyperref}
\usepackage{dcolumn}
\usepackage{bm}
\usepackage{subfigure}
\usepackage{amsthm}
\usepackage[english]{babel}
\usepackage{titlesec}
\usepackage{amsmath}
\usepackage{amsfonts}
\usepackage{amssymb}
\usepackage{mathrsfs}
\usepackage{lipsum}
\usepackage[pdftex]{color,graphicx}

\usepackage[percent]{overpic}
\usepackage[export]{adjustbox}

\usepackage{xcolor}

\begin{document}

\title{Fidelity plateaux from correlated noise in isolated few-level quantum systems}
\author{Scott R. Taylor}
\affiliation{SUPA, School of Physics and Astronomy, University of St Andrews, North Haugh, St Andrews, Fife KY16 9SS, United Kingdom}
\affiliation{Abdus Salam International Centre for Theoretical Physics, Strada Costiera 11, 34151 Trieste, Italy}
\author{Chris A. Hooley}
\affiliation{SUPA, School of Physics and Astronomy, University of St Andrews, North Haugh, St Andrews, Fife KY16 9SS, United Kingdom}
\date{Sunday 24th May 2020}

\begin{abstract}
We show that, in an isolated two-level quantum system described by a time-dependent Hamiltonian, correlated noise in the Hamiltonian's parameters can lead to an arbitrarily long plateau in the state-preparation fidelity as a function of elapsed time.
We explain the formation of this plateau using the Bloch-sphere representation, deriving analytical expressions for its start and end times and its height.
We also briefly discuss the extent to which this phenomenon is expected to be visible in more general quantum systems with $N>2$ levels.
\end{abstract}
\maketitle
 
\section{Introduction}
Quantum systems described by Hamiltonians that include random elements have been studied for many decades.  The majority of that work has focused on the case where the Hamiltonian is independent of time.  One example is random-matrix theory \cite{Guhr1997,OxfordRMT,RandomMatrices}, where each entry in the Hamiltonian is drawn independently from a given probability distribution, subject only to global symmetry constraints on the resulting matrix.  Another is the study of quenched disorder in condensed matter systems \cite{Anderson1958,Abrahams1979,Belitz1994,Evers2008}, where typically a random part (e.g.\ on-site energies) is added to an otherwise non-random Hamiltonian (e.g.\ a $d$-dimensional tight-binding model).  A modern variant of the latter is many-body localization \cite{Basko2006,Nandkishore2015}, where the non-random Hamiltonian to which the disorder is added includes interparticle interactions.

Another area that has been studied since the early days of quantum mechanics is Hamiltonians that are wholly or partly time-dependent.  Here the canonical example is the Landau-Zener problem \cite{Landau1932,Zener1932,Majorana1932,Stueckelberg1932}, where one studies the probability of transfer from the ground to the excited state as a two-level system is driven non-adiabatically through an anticrossing in its spectrum.  There are many variants on this problem, including some exactly solvable cases \cite{Rosen1932,Demkov1964,Demkov1968,Hioe1985,Ostrovsky1997}.  Recent progress in experimental cold-atom physics \cite{Bloch2008} has led to increased interest in problems of driven isolated quantum systems, though in fact these also arise naturally in several other fields of physics, notably nuclear magnetic resonance \cite{Hore2015}.  Furthermore, the drive to implement quantum information processing has led to a concerted effort to produce good approximations to driven isolated quantum systems in such systems as cold-atom optical lattices \cite{Gross2017}, trapped ions \cite{Blatt2012}, impurities in semiconductors \cite{Awschalom2013}, NV centers in diamond \cite{Sipahigil2012}, superconducting circuits \cite{Devoret2013}, and more.

Systems that possess both of these features, i.e.\ for which the Hamiltonian contains both explicit time-dependence and randomness, exhibit a rich range of physical phenomena.  Again there are subclassifications, depending on whether or not the division between the non-random (`clean') and random (`disordered') parts of the Hamiltonian lines up with the division between the time-independent and time-dependent parts.  A particular case of recent interest is the clean monochromatic driving of a disordered system, which is a particular example of the Floquet problem \cite{Eckardt2017}.

In the case where the time-dependent term is disordered, one might suppose that the strong-disorder limit is simple, corresponding in some sense to `full randomization' of the state over time.  In this paper, we present a counter-example to that supposition:\ a two-level system in which the application of strong time-dependent disorder (`noise') leads to a long-lived and non-trivial metastable state.  In section \ref{s:model} we give the Hamiltonian of the model in question, discuss its potential physical realizations, and introduce the fidelity, the main observable of interest to us.  In section \ref{s:plateau}, we analyze the strong-noise limit of our model, predicting a long-lived plateau at a non-trivial value of the fidelity which we calculate.  In section \ref{s:numerics}, we present numerical evidence of the existence of this plateau, and in section \ref{s:conditions} we derive expressions for the start- and end-times of the plateau which we again confirm numerically.  In section \ref{s:outlook}, we summarize our findings, and discuss to what extent the emerging picture is valid for more general quantum systems subjected to strong correlated noise.

\section{The model} \label{s:model}
Our model starts from an ideal noise-free state-preparation process for a two-level quantum system.  This process is described by a time-dependent Hamiltonian $\mathcal{H}_0(t)$ that evolves from an initial value at time $t=0$ to a final one at time $t=T$; we refer to $T$ as the exposure time. We define the instantaneous eigenstates of $\mathcal{H}_0(t)$ as follows:
\begin{eqnarray}
\mathcal{H}_0(t) | g_0(t) \rangle & = & E^0_g(t) | g_0(t) \rangle; \\
\mathcal{H}_0(t) | e_0(t) \rangle & = & E^0_e(t) | e_0(t) \rangle,
\end{eqnarray}
where `$g$' stands for `ground', and `$e$' for `excited'.  We start the system at $t=0$ in the instantaneous ground state of the Hamiltonian $\mathcal{H}_0(0)$, i.e.\ $| \psi(0) \rangle = | g_0(0) \rangle$, and we define the fidelity, $\mathscr{F}(t)$, as the probability of finding the system in the instantaneous ground state of $\mathcal{H}_0(t)$ at time $t$:
\begin{equation}
\label{eq:fidelity}
\mathscr{F}(t) = \big| \langle g_0 (t) | \psi (t) \rangle \big|^2.
\end{equation}

Our measure of adiabaticity is the probability of finding the system in its ground state at the end of the process, $\mathscr{F}(T)$.
Of course this probability will not in general be unity, because of the possibility of non-adiabatic transitions out of the ground state due to the finite rate of change of $\mathcal{H}_0(t)$.  
In a noise-free system, however, provided the Hamiltonian changes smoothly and the energy gap above the ground state remains finite throughout, $\mathscr{F}(T) \to 1$ as $T$ becomes large \cite{Messiah}.

In a system with fast noise this is not the case, since the eigenbasis fluctuates rapidly even for large $T$.  To include such noise, we add to our Hamiltonian a rapidly fluctuating term $\mathcal{H}_1(t)$:
\begin{equation}
\mathcal{H}(t) = \mathcal{H}_0(t) + \mathcal{H}_1(t). \label{noisehamgen}
\end{equation}
In general the real and imaginary parts of each matrix element of $\mathcal{H}_1(t)$ might have independent and uncorrelated time-dependence.  However, in this paper we shall concentrate on the case where a single fast time-dependence $\eta(t)$ is common to all of the matrix elements, though potentially multiplied by a slow variation that is matrix-element-dependent.  This kind of noise would be experienced, for example, by a spin-1/2 particle traveling through a long and slightly curved solenoid carrying a rapidly fluctuating current:\ in this situation, the magnitude of the applied magnetic field would vary rapidly, superposed on a much slower variation of its direction.  As we shall show, such `correlated noise' shows some counter-intuitive features in the time-dependence of the fidelity.

The Hamiltonian (\ref{noisehamgen}) in the case of perfectly correlated noise may be written
\begin{equation}
\label{eq:H}
\mathcal{H}(t) = \mathcal{H}_0(t) + \epsilon \,\eta(t) \, \mathcal{H}_n(t),
\end{equation}
where $\mathcal{H}_0$ and $\mathcal{H}_n$ are operators that change smoothly and deterministically in time, $\eta(t)$ is rapidly fluctuating noise, and $\epsilon$ is a parameter that quantifies the strength of the noise.
We take the noise to be a Gaussian-distributed stochastic variable with zero mean, $\overline{\eta(t)} = 0$, and a two-time correlator
\begin{equation}
\label{eq:correlator}
\overline{\eta (t) \eta (t + \tau)} = \delta (\tau),
\end{equation}
where the bar denotes an average over realizations of the noise.  The delta function should be considered as an approximation, valid in the limit where the correlation time of the noise is much shorter than any other timescale in the system \cite{Gard, Gardiner}.

We have already introduced notation for the instantaneous eigenstates and eigenvalues of $\mathcal{H}_0(t)$; analogously, we denote the eigenstates and eigenvalues of $\mathcal{H}_n(t)$ by
\begin{eqnarray}
\mathcal{H}_n(t) | g_n(t) \rangle & = & E^n_g(t) | g_n(t) \rangle; \\
\mathcal{H}_n(t) | e_n(t) \rangle & = & E^n_e(t) | e_n(t) \rangle.
\end{eqnarray}
We will initialize the system and measure the fidelity with reference to the ground state of the clean part of the Hamiltonian $| g_0(t) \rangle$.

\section{The fidelity plateau} \label{s:plateau}
When the noise is strong, the Hamiltonian \eqref{eq:H} will be dominated by the contribution from the noise part, i.e.
\begin{equation}
\label{hamsn}
\mathcal{H}(t) \approx \epsilon\,\eta(t)\,\mathcal{H}_n (t).
\end{equation}
In this limit, we should therefore describe the system in the basis $\{ |g_n(t)\rangle, |e_n(t)\rangle\}$, the instantaneous eigenbasis of $\mathcal{H}_n(t)$.  
If the time-evolution of $\mathcal{H}_n(t)$ is slow enough that the system adiabatically follows the eigenstates of $\mathcal{H}_n(t)$ then the Hamiltonian (\ref{hamsn}) causes pure dephasing, i.e.\ the relative populations of $|g_n(t)\rangle$ and $|e_n(t)\rangle$ become independent of time.  This, as we shall show, gives rise to a plateau in the disorder-averaged fidelity $\overline{\mathscr{F}(T)}$.

This `pure dephasing' time-evolution is given by
\begin{equation}
\label{eq:following}
| \psi (t) \rangle = \sum_{\mu \in  \{ g, e \}} c_{\mu}
				\exp \left(- \frac{i}{\hbar} \int_0^t \mathrm{d}t' \epsilon \, \eta(t') \, E^n_{\mu}(t')\right)
				| \mu_n (t) \rangle,
\end{equation}
where $c_{\mu} = \langle \mu_n (0) | \psi (0) \rangle = \langle \mu_n (0) | g_0 (0) \rangle$.
When the relative phase of the two components of $| \psi (t) \rangle$ is fully randomized, the average over disorder realizations gives a fidelity of
\begin{eqnarray}
\label{eq:F0first}
\mathscr{F}_0(t) & = & \big| \, \langle g_n (0) | g_0(0) \rangle \,
				\langle g_0(t) | g_n (t) \rangle \, \big|^2 \nonumber \\
& & \quad + \,\big| \, \langle e_n (0) | g_0(0) \rangle \, \langle g_0(t) | e_n (t) \rangle \, \big|^2,
\end{eqnarray}
which depends only on the instantaneous properties of $\mathcal{H}_0(0)$, $\mathcal{H}_0(t)$, $\mathcal{H}_n(0)$, and $\mathcal{H}_n(t)$, not on $T$.
Thus, counterintuitively, strong noise does not fully randomize the system's state, instead distributing it uniformly over a manifold of lower dimension than the full state space.

\begin{figure}
\begin{center}
\includegraphics[width=\columnwidth]{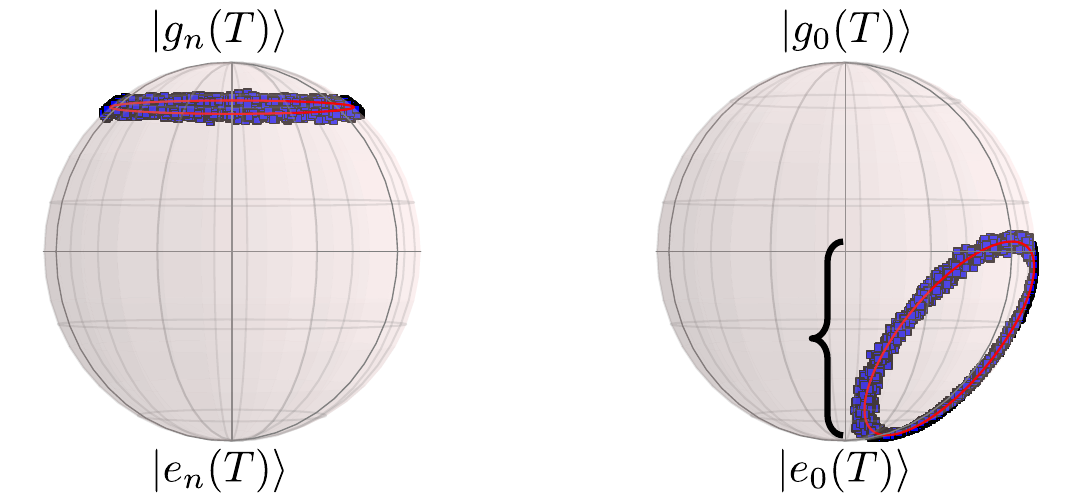}
\caption{
	The distribution of final states on the Bloch sphere in the `pure dephasing' limit. The sphere on the left 
	represents the state of the system in the $\mathcal{H}_n (t)$ eigenbasis; the sphere on the right in the 
	$\mathcal{H}_0 (t)$ eigenbasis.  On the right the distribution of final-state fidelities (i.e.\ the range of 
	$z$-coordinates) is indicated by a left-brace.
	}
\label{fig:Bloch}
\end{center}
\end{figure}

We may visualize this by comparing the evolution on the Bloch spheres defined by the instantaneous eigenstates of $\mathcal{H}_0(t)$ and $\mathcal{H}_n(t)$.
On the $\mathcal{H}_n(t)$ Bloch sphere the initial state is located at a point on the sphere's surface at some angle $\theta_0$ to the $|g_n(t)\rangle$ pole.
Pure-dephasing behavior then corresponds to the state performing a random walk around the line of latitude $\theta=\theta_0$.
Thus at $t=T$ the ensemble of final states is distributed on a ring at a constant angle $\theta_0$ from the $|g_n(T)\rangle$ pole, as shown on the left of Fig.~\ref{fig:Bloch}.

The unitary transformation relating the eigenstates of $\mathcal{H}_0(T)$ to those of $\mathcal{H}_n(T)$ corresponds to a rotation of the Bloch sphere.
This rotation preserves the form of the ring of final states, but changes the angular position of its center, as shown on the right of Fig.~\ref{fig:Bloch}.

The fidelity is equal to half of the $z$-coordinate of the state on the Bloch sphere measured from the $|e_0(T)\rangle$ pole.  This results in a distribution of $\mathscr{F}(T)$, the average of which (equal to $\mathscr{F}_0(T)$) corresponds to the $z$-coordinate of the center of the ring as described above.
The distribution of fidelities corresponding to this ring is:
\begin{equation}
\label{eq:Fdistribution}
f (x) = \left\{
\begin{array}{lll}
\displaystyle 
\frac{1}{\pi} \left[ \mathscr{F}_1^2 - \left( x - \mathscr{F}_0 \right)^2 \right]^{-1/2} & \,\,\,\, & \left\vert x - \mathscr{F}_0 \right\vert \leqslant \left\vert \mathscr{F}_1 \right\vert; \\
& & \\
0 & & \mbox{otherwise,}
\end{array}
\right.	
\end{equation}
where $\mathscr{F}_1 = 2 \big| \big\langle g_n^{(i)} \big| g_0^{(i)} \big\rangle  \big\langle g_0^{(f)} \big| g_n^{(f)} \big\rangle  \big\langle e_n^{(i)} \big| g_0^{(i)} \big\rangle \big\langle g_0^{(f)} \big| e_n^{(f)} \big\rangle \big|$, with the superscripts $(i)$ and $(f)$ signifying the state evaluated at $t=0$ and $t=T$ respectively.

\section{Numerical demonstration of the fidelity plateau} \label{s:numerics}
\begin{figure}
\begin{center}
\includegraphics[width=\columnwidth]{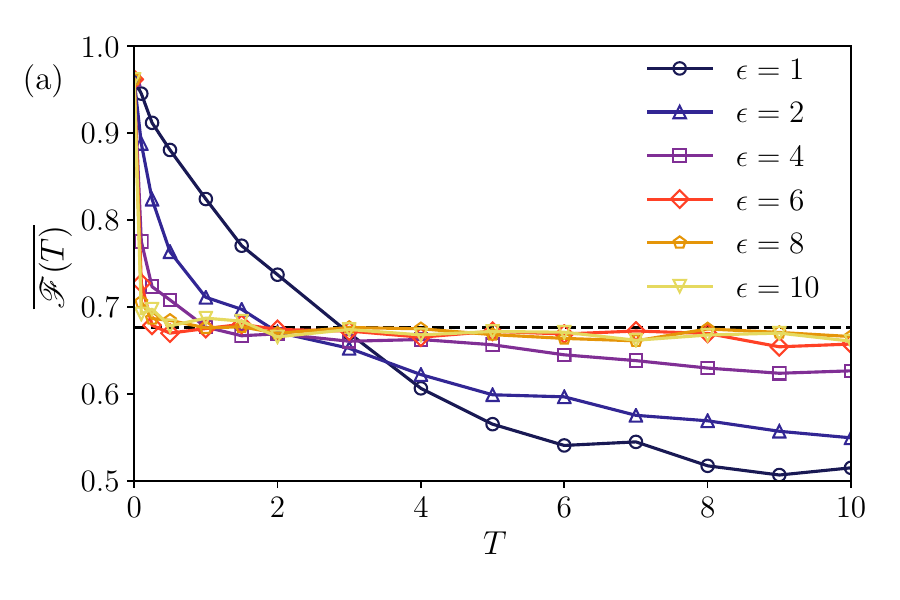}\\
\includegraphics[width=\columnwidth]{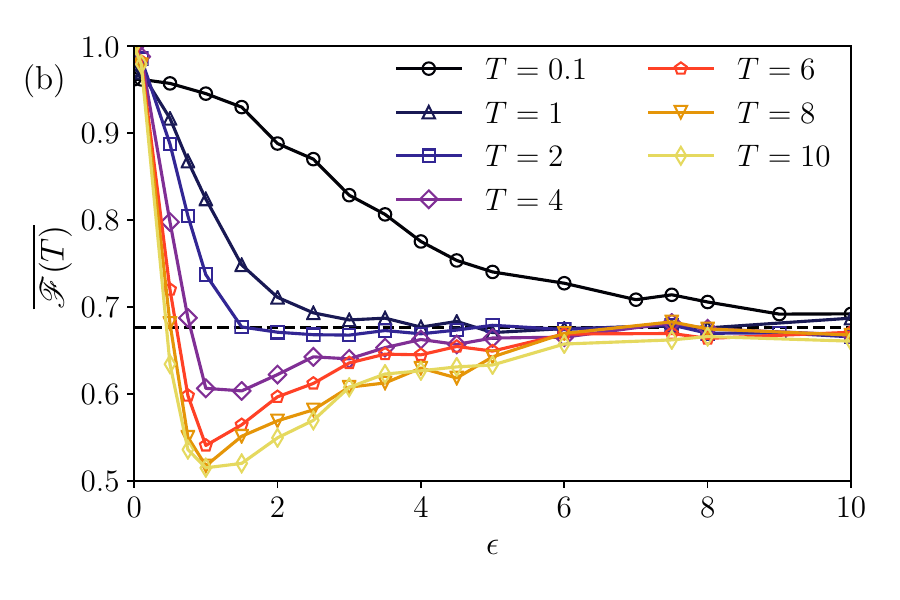}
\caption{
	Disorder-averaged state-preparation fidelities $\overline{\mathscr{F}(T)}$, determined by numerical time-evolution according to the Hamiltonian (\ref{eq:Hex}).  The plateau value 
	$\mathscr{F}_0(T)$ --- see equation \eqref{eq:F0second} --- is indicated by the black dashed lines.
	(a)  $\overline{\mathscr{F}(T)}$ plotted as a function of exposure time $T$ for a range of noise 
	strengths $\epsilon$. For strong noise, the fidelity exhibits a plateau at $\mathscr{F}_0(T)$ rather than 
	decaying directly to $1/2$. Increasing the noise strength causes the plateau to
	begin earlier and to end later. For $T \to 0$ the fidelity tends to $| \langle g_0 (0) | g_0 (T) \rangle |^2 \approx 0.962$.
	(b) $\overline{\mathscr{F}(T)}$ plotted as a function of noise strength for a range of 
	exposure times. For strong noise the fidelities tend to $\mathscr{F}_0(T)$ for all exposure times shown.
	}
\label{fig:plateau}
\end{center}
\end{figure}
We now demonstrate the predicted behavior in a simple model by numerically time-evolving the system using the Heun algorithm \cite{Gard}.
We choose a Hamiltonian where $\mathcal{H}_0(t)$ and $\mathcal{H}_n(t)$ both rotate at constant rates with constant energy gaps:
\begin{equation}
\begin{aligned}
\label{eq:Hex}
\mathcal{H} (t) & = \overbrace{\hbar \omega_0 \left[ \cos \left( \vartheta_0 (t) \right) \sigma^z + \sin \left( \vartheta_0 (t) \right) \sigma^x \right]}^{\mathcal{H}_0(t)} \\
& \quad + \, \epsilon \, \eta (t) \, \underbrace{\hbar \omega_n \left[ \cos \left( \vartheta_n (t) \right) \sigma^z + \sin \left( \vartheta_n (t) \right) \sigma^x \right]}_{\mathcal{H}_n(t)},
\end{aligned}
\end{equation}
where
\begin{equation}
\vartheta_{\alpha} (t) = \vartheta_{\alpha}(0) + \left(  \vartheta_{\alpha}(T) -  \vartheta_{\alpha}(0) \right) \frac{t}{T}, \quad \alpha \in \{0,n\},
\label{eq:theta}
\end{equation}
and $\sigma^{\nu}$ with $\nu \in \{ x, z \}$ are Pauli matrices.
This choice of Hamiltonian results in simple analytical expressions.  However, the analysis presented in sections \ref{s:plateau} and \ref{s:conditions} is valid for $\mathcal{H}_0(t)$ and $\mathcal{H}_n(t)$ with arbitrary (smooth) time-dependences and including $\sigma^y$ terms. As such, the qualitative features of our numerical results should be common to any two-level system with a Hamiltonian of the form \eqref{eq:H}.

\begin{figure}
\begin{center}
\includegraphics[width=\columnwidth]{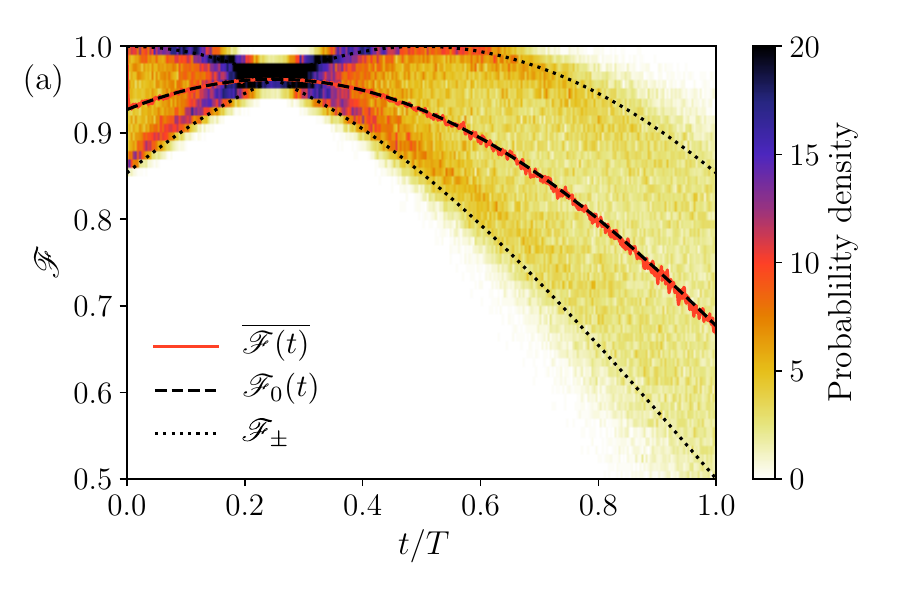}\\
\includegraphics[width=\columnwidth]{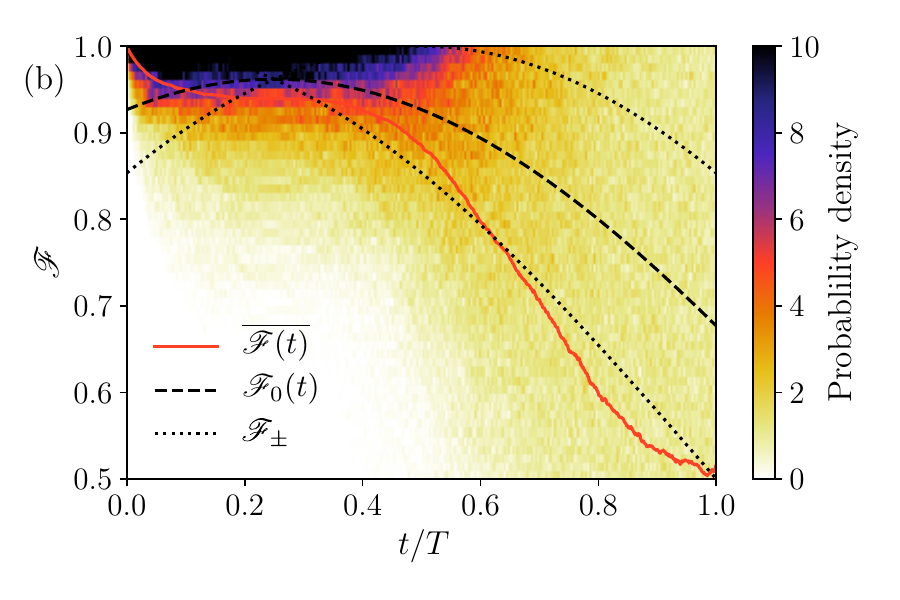}
\caption{Time-resolved histograms of the instantaneous fidelity, determined by numerical time-evolution according to the Hamiltonian (\ref{eq:Hex}).  The solid red line shows the disorder-averaged fidelity $\overline{\mathscr{F}(T)}$.  For comparison, we also show the predicted average $\mathscr{F}_0(t)$ --- see equation (\ref{eq:F0second}) --- by the dashed black line, and the predicted bounds --- see equation (\ref{eq:mylims}) --- by the dotted black lines.
	      (a) In the strong-noise limit (here $T = 4$, $\epsilon = 5 \sqrt{T}$) the distribution of fidelities stays within the 
	      predicted bounds, and the average is equal to $\mathscr{F}_0(t)$.
	      (b) Outside the strong-noise limit (here $T = 10$, $\epsilon = \sqrt{T/10}$) the fidelities do not show this 
	      behavior.
	      }
\label{fig:fidelities}
\end{center}
\end{figure}

In this model the instantaneous eigenstates of $\mathcal{H}_{\alpha} (t)$ are:
\begin{eqnarray}
\big| g_{\alpha} (t) \big\rangle = \sin \frac{\vartheta_{\alpha}(t)}{2} \big| {\uparrow} \big\rangle -  \cos \frac{\vartheta_{\alpha}(t)}{2} \big| {\downarrow} \big\rangle; \\
\big| e_{\alpha} (t) \big\rangle = \cos \frac{\vartheta_{\alpha}(t)}{2} \big| {\uparrow} \big\rangle +  \sin \frac{\vartheta_{\alpha}(t)}{2} \big| {\downarrow} \big\rangle,
\end{eqnarray}
and the corresponding eigenvalues are $\mp \hbar \omega_{\alpha}$.
In this model, the fidelity of the plateau is given by:
\begin{equation}
\begin{aligned}
\label{eq:F0second}
\mathscr{F}_0(t) = & \cos^2 \left( \frac{\delta\vartheta (0)}{2} \right) \, \cos^2 \left( \frac{\delta\vartheta (t)}{2} \right) \\
& \quad + \, \sin^2 \left( \frac{\delta\vartheta (0)}{2} \right) \, \sin^2 \left( \frac{\delta\vartheta (t)}{2} \right)
\end{aligned}
\end{equation}
where $\delta\vartheta (t) = \vartheta_n (t) - \vartheta_0 (t)$.
According to \eqref{eq:following}, for sufficiently large $T$ the fidelity of any individual noise realization must stay within the bounds:
\begin{equation}
\begin{aligned}
\label{eq:mylims}
\mathscr{F}_{\pm} (t) = & \left( \cos \left( \frac{\delta\vartheta (0)}{2} \right) \, \cos \left( \frac{\delta\vartheta(t)}{2} \right) \right. \\
& \quad \pm \, \left. \sin \left( \frac{\delta\vartheta(0)}{2} \right) \, \sin \left( \frac{\delta\vartheta(t)}{2} \right) \right)^2,
\end{aligned}
\end{equation}
which correspond to the top and bottom of the ring on the right of Fig.~\ref{fig:Bloch}.
For these numerical tests we use the following parameters: $\omega_0=\omega_n=1$, $\vartheta_0 (0)=\pi/4$, $\vartheta_0 (T)=\pi/8$, $\vartheta_n (0)=\pi/8$, $\vartheta_n (T)=\pi/2$.
This results in a plateau fidelity of $\mathscr{F}_0(T) \approx 0.677$, and we note that $\vartheta_0 (T/4) = \vartheta_n (T/4)$, so at this time the limits $\mathscr{F}_{\pm}(t)$ close to a point.
The results are shown in Figs.~\ref{fig:plateau} and \ref{fig:fidelities}, where data has been gathered over $N$ disorder realizations such that the standard error $\sigma / \sqrt{N}$ of $\mathscr{F}(T)$ is at most $1 \%$ of the mean.

Fig.~\ref{fig:plateau}(a) shows the disorder-averaged fidelity as a function of exposure time for various noise strengths.  It decays from $\mathscr{F}(0) = |\langle g_0(T) | g_0(0) \rangle|^2 \approx 0.962$ to the value $\mathscr{F}_0(T)$, indicated by the dashed line.
For strong noise, as predicted, it `sticks' there, forming a long plateau; for weak noise, it almost immediately resumes decaying towards $1/2$, the value na\"\i vely expected for a two-state system.

Fig.~\ref{fig:plateau}(b) shows the disorder-averaged fidelity $\overline{\mathscr{F}(T)}$ as a function of noise strength for various exposure times.
The predicted value of the fidelity plateau $\mathscr{F}_0$ is shown by the dashed line.
For all exposure times shown, the average fidelity tends to $\mathscr{F}_0$ as the noise becomes strong.
The curves depart from the strong-noise prediction as the noise strength $\epsilon$ is reduced.
This reflects the counter-intuitive fact that weaker noise is better able to spread the ensemble of states across the entire Bloch sphere.

Fig.~\ref{fig:fidelities} shows histograms (vertical slices) of $\mathscr{F}(t)$ as a function of time for systems both (a) in the strong-noise limit (for $T = 4$ and $\epsilon = 5 \sqrt{T}$) and (b) outside the strong-noise limit (for $T = 10$ and $\epsilon = \sqrt{T/10}$).
The dashed black line shows $\mathscr{F}_0(t)$ (the center of the ring in the right-hand panel of Fig.~\ref{fig:Bloch}), the dotted black lines show the upper and lower bounds on $\mathscr{F}(t)$ (the top and bottom of the ring in the right-hand panel of Fig.~\ref{fig:Bloch}), and the solid red line shows $\overline{\mathscr{F}(t)}$.
In the strong-noise limit the average fidelity falls exactly on the $\mathscr{F}_0(t)$ curve, while the full distribution is well contained by the predicted bounds even when they close to a point at $t=T/4$.
In panel (b) this is not the case, and the fidelity becomes uniformly distributed between 0 and 1.

\section{When does the plateau occur?} \label{s:conditions}
Three conditions must be satisfied for the fidelity plateau to occur.  They are most easily understood in terms of the equation of motion for $c_g(t)$ and $c_e(t)$, the decomposition of the system's state in terms of the instantaneous eigenbasis of $\mathcal{H}_n$:
\begin{eqnarray}
i \hbar
 \begin{pmatrix}
  {\dot c}_g \\ {\dot c}_e
 \end{pmatrix} & = & \Bigg[ 
   \begin{pmatrix}
  \langle g_n | \mathcal{H}_0 | g_n \rangle & \langle g_n | \mathcal{H}_0 | e_n \rangle \\
  \langle e_n | \mathcal{H}_0 | g_n \rangle & \langle e_n | \mathcal{H}_0 | e_n \rangle
 \end{pmatrix} \nonumber \\
 & & \!\!\!\!\!\!\!\!\!\!\!\!\!\!\!\!\!\!\!\!\!\!\!\!\! -\,i \hbar
 \begin{pmatrix}
  \langle g_n | \dot{g}_n \rangle & \langle g_n | \dot{e}_n \rangle \\
   \langle e_n | \dot{g}_n \rangle & \langle e_n | \dot{e}_n \rangle
 \end{pmatrix} 
 + \epsilon \,\eta
  \begin{pmatrix}
  E^n_g & 0\\
  0 & E^n_e
 \end{pmatrix} \Bigg]
  \begin{pmatrix}
  c_g \\ c_e
 \end{pmatrix}, \;\;\;\;\;\;
 \label{eq:EOM}
\end{eqnarray}
where the dots denote time-derivatives.
The first condition is that transitions due to $\left[ \mathcal{H}_0, \mathcal{H}_n \right] \neq 0$ must be negligible, i.e.\ the first term in the square brackets must be small compared to the third.  The second condition is that non-adiabatic transitions between the instantaneous eigenstates of $\mathcal{H}_n$ must also be negligible, i.e.\ the second term must be small compared to the third.  The third condition is that the third term must be strong enough to completely dephase the components of the initial state by the end of the process.

\begin{figure}
\begin{center}
\includegraphics[width=\columnwidth]{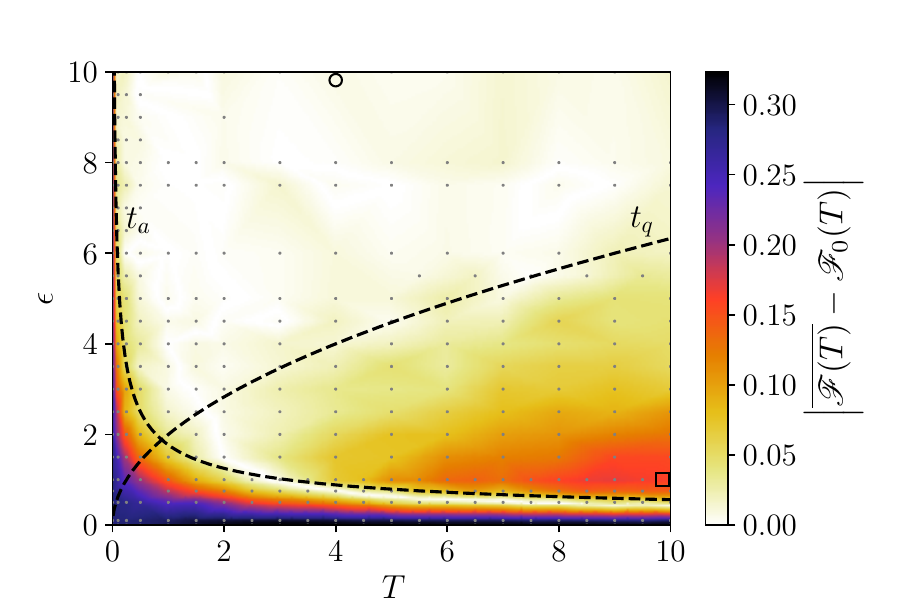}
\caption{$\big| \overline{\mathscr{F}(T)} - \mathscr{F}_0 \big|$, the absolute deviation of the disorder-averaged state-preparation fidelity from the plateau value,
	        as a function of exposure time $T$ and noise
	        strength $\epsilon$.
	        The black dashed lines show the boundaries of the plateau, as predicted by \eqref{eq:tq}
	        and \eqref{eq:ta}. The boundaries plotted here show the plateau at $t_a < T < t_q$, where
	        $t_a = (6 \hbar D_n)^2 / (\epsilon \, (\Delta E_n)^2)^2$ and
	        $t_q = (\epsilon \, \Delta E_n)^2 / (2 \, \Delta E_0)^2$, with the numerical coefficients chosen by eye.
		We average over a number of disorder realizations such that the standard error $\sigma / \sqrt{N}$ is at
		most $1.5 \%$ of the mean. The light gray markers indicate data points, and the color map is determined
		by a linear interpolation between these points.
		The black circle and square indicate the parameter values used in Figs.~\ref{fig:fidelities}(a) and \ref{fig:fidelities}(b) respectively.
		}
\label{fig:scaling}
\end{center}
\end{figure}
To determine when the first condition is satisfied, we rescale the time variable:\ $t = s \, T$ with $s \in [0,1]$.
For the first term in the square brackets in (\ref{eq:EOM}), this is equivalent to rescaling the matrix elements by a factor $T$.
However, the third term must be treated more carefully.  Examining the correlation function of the noise,
\begin{equation}
\label{eq:corr}
\overline{\eta(t)\eta(t')} = \delta(t-t') = \delta \left(T[s-s']\right) = T^{-1} \delta(s-s'),
\end{equation}
we see that the appropriate rescaling is $\eta(t) \to T^{-1/2} \eta(s)$.
Thus the Schr\"odinger equation becomes:
\begin{equation}
\label{eq:SEscaled}
i \hbar \partial_s | \psi (s) \rangle  = \left[ T \mathcal{H}_0(s) + \epsilon \sqrt{T} \, \eta(s) \, \mathcal{H}_n(s) \right] 
| \psi (s) \rangle.
\end{equation}
This shows that the `strength' of the noise depends on how long the system is exposed to it.
The influence of the deterministic part becomes comparable to that of the noise when the two terms on the right-hand side of \eqref{eq:SEscaled} are of similar magnitude:
$T \, \Delta E_0 \sim \epsilon \sqrt{T} \, \Delta E_n$,
where $\Delta E_{\alpha}$ is the typical energy difference between the eigenstates of $\mathcal{H}_{\alpha}$.
Thus we may neglect transitions due to $\mathcal{H}_0$ provided that
\begin{equation}
\label{eq:tq}
T \ll t_{q} \sim \left( \epsilon \frac{\Delta E_n}{\Delta E_0} \right)^2,
\end{equation}
while for very large $T$ the state becomes completely randomized \cite{Pokrovsky2003}.

The second condition requires that the third term on the right-hand side of \eqref{eq:EOM} is much stronger than the second.
Rescaling time as before, we find that non-adiabatic effects can no longer be neglected when
$\epsilon \sqrt{T} \, \Delta E_n \sim \hbar \, D_n \, \Delta E_n^{-1},$
where $D_n$ is the typical magnitude of $\langle g_n (s) | \partial_s \mathcal{H}_n (s) | e_n (s) \rangle$.
Thus, as expected, non-adiabatic effects become weaker for processes performed over a longer time, becoming negligible when
\begin{equation}
\label{eq:ta}
T \gg t_a \sim \left( \frac{\hbar \, D_n}{\epsilon \, (\Delta E_n)^2} \right)^2.
\end{equation}

Assuming that the first two conditions are satisfied, only the third term on the right-hand side of (\ref{eq:EOM}) need be retained.  Thus, from equation \eqref{eq:following}, we see that the relative phase at time $T$ is given by $\Delta \phi =  \frac{1}{\hbar} \int_0^{T} \mathrm{d}t \, \epsilon \, \eta(t) \, \Delta E_n (t)$,
where $\Delta E_n (t) = E^n_e(t) - E^n_g(t)$, the instantaneous gap between the eigenenergies of $\mathcal{H}_n$.
It follows that
$\overline{\Delta \phi} = 0$ and $\overline{(\Delta \phi)^2} = \epsilon^2 \hbar^{-2} \, \delta E_n^2 \, T$,
where $\delta E_n^2 = T^{-1}\int_0^T \mathrm{d}t \left[ \Delta E_n(t) \right]^2$.
The distribution of $\Delta \phi$ becomes approximately uniform when $\overline{(\Delta \phi)^2}$ is of order one, and thus the phase is randomized provided that
\begin{equation}
\label{eq:tphi}
T \gg t_{\phi} \sim \frac{\hbar^2}{\epsilon^2 \, \delta E_n^2}.
\end{equation}
We therefore expect to observe the plateau when the Hamiltonians $\mathcal{H}_0$ and $\mathcal{H}_n$, the noise strength $\epsilon$, and the exposure time $T$ are such that the following condition is satisfied:
\begin{equation}
t_{\phi}, \, t_a \ll T \ll t_q.
\label{eq:valcon}
\end{equation}

Fig.~\ref{fig:scaling} shows $\big\vert \overline{\mathscr{F}(T)} - \mathscr{F}_0 \big\vert$, the deviation of the numerically determined fidelity from the predicted plateau value for the model described by \eqref{eq:Hex}, as a function of noise strength $\epsilon$ and exposure time $T$.
The light gray markers indicate data points, and the color map is determined by linearly interpolating between these points.
The dashed black lines show the timescales \eqref{eq:tq} and \eqref{eq:ta}, and we see that the fidelity attains its plateau value everywhere between these two curves on the side where $t_q > t_a$, exactly as predicted by \eqref{eq:valcon}.
For the $\mathcal{H}_n$ studied in section~\ref{s:numerics} one finds that $D_n = \vartheta_n(T) - \vartheta_n(0) = 3 \pi / 8$.
The curve $\eqref{eq:tphi}$ is not marked on the graph as it has the same functional form as $t_a$, and for this system we find that $t_\phi < t_a$ for all parameter values used in generating Figs.~\ref{fig:plateau}-\ref{fig:scaling}.  In particular, this can be seen from the rapid collapse of $\overline{\mathscr{F}(t)}$ onto $\mathscr{F}_0(t)$ in Fig.~\ref{fig:fidelities}(a).

\section{Summary and outlook} \label{s:outlook}
In this paper we have described in detail, both analytically and numerically, a particular model of time-dependent state-preparation for an isolated two-level quantum system subjected to strong noise.  We have shown that, for a broad range of parameters, this results in a long-lived metastable state which manifests itself as a plateau in the state-preparation fidelity as a function of time.  The key observation is that, on the appropriate Bloch sphere, strong correlated noise drives the system on a single orbit, leading to partial but not total randomization of the state.

How does this generalize to the case of an $N$-level system?  To answer this question, let us consider the above two-level case from the group-theoretic point of view.  A generic two-level Hamiltonian can be expressed as a linear combination of the three Pauli matrices, which are the generators of the group SU(2), plus the identity.  The Lie algebra of SU(2) has a one-dimensional Cartan subalgebra formed by just one of its generators, e.g.\ $\sigma^z$.  Fast noise coupled to only this generator cannot fully randomize the state, since it can only rotate the spin around a single axis.  As we have shown, this phenomenon survives over a broad range of times even if the direction of that axis is allowed to evolve slowly in time.

In the $N$-level case, the group is SU($N$) rather than SU(2).  SU($N$) has $N^2-1$ generators, of which $N-1$ form a Cartan subalgebra.  We may take these $N-1$ generators to be represented by traceless diagonal matrices.  Even if each such generator were multiplied by an independent fast noise function $\eta_\alpha(t)$, $\alpha = 1,2, \ldots,N-1$, the sum of such terms could not fully randomize the state, since the overall fast-noise Hamiltonian would still be diagonal:\ it could induce only relative dephasing of the different components of the wave function in the Cartan basis.  Presumably, as in the SU(2) case, this would remain true over a broad range of times even if the matrices multiplying the fast noise functions $\eta_\alpha(t)$ evolved slowly in time.  However, this slow evolution would have to preserve the mutually commuting nature of the $N-1$ generators in the Cartan subalgebra, and would thus be less general than in the SU(2) case.  Given this caveat, however, we expect that the essential physics described here will extend to the $N$-level system, though presumably with some changes to the adiabaticity condition (\ref{eq:ta}); we defer the detailed investigation of these to a forthcoming work \cite{inprep}.

\textit{Acknowledgments.} SRT acknowledges financial support from the CM-CDT under EPSRC (UK) grant number EP/L015110/1.  CAH acknowledges financial support from the TOPNES programme under EPSRC (UK) grant number EP/I031014/1.  This research was supported in part by the National Science Foundation under Grant No.\ NSF PHY-1125915.  Part of this work was performed at the Aspen Center for Physics, which is supported by National Science Foundation grant PHY-1066293.  Parts of this work were carried out at the Max Planck Institute for the Physics of Complex Systems in Dresden and at Rice University in Houston, Texas:\ CAH gratefully acknowledges their hospitality.

\bibliographystyle{apsrev4-1}
\bibliography{references}
\end{document}